\documentclass[12pt,preprint]{aastex} 
\usepackage{natbib}
\newcommand{\degree}{\hbox{$^\circ$}}
\newcommand{\ergseccm}{erg\,sec$^{-1}$\,cm$^{-2}$}
\newcommand{\etal}{et\,al.}
\newcommand{\halpha}{H$\alpha$}
\newcommand{\hbeta}{H$\beta$}

\newcommand{\gsim}{\raise0.3ex\hbox{$>$}\kern-0.75em{\lower0.65ex\hbox{$\sim$}}}
\newcommand{\kms}{km\,s$^{-1}$}

\newcommand{\lsim}{\raise0.3ex\hbox{$<$}\kern-0.75em{\lower0.65ex\hbox{$\sim$}}}

\newcommand{\mjpbeam}{\,\,mJy\,Beam$^{-1}$}
\newcommand{\ujpbeam}{\,\,$\mu$Jy\,Beam$^{-1}$}
\newcommand{\msun}{M$_{\odot}$}

\newcommand{\HI}{H~{\sc i}}
\newcommand{\HII}{H~{\sc ii}}

\begin{document}     
\slugcomment{Accepted for publication in the Astrophysical Journal}
\title{The Complex Neutral Gas Dynamics Of The Dwarf Starburst Galaxy NGC 625}
\author{John M. Cannon}
\affil{Department of Astronomy, University of Minnesota,\\ 116 Church St. 
S.E., Minneapolis, MN 55455}
\email{cannon@astro.umn.edu}
\author{N.\,M. McClure-Griffiths\altaffilmark{1}}
\affil{Australia Telescope National Facility, CSIRO,\\
P.O. Box 76, Epping, NSW 1710, Australia}
\email{nmmclure@atnf.csiro.au}
\altaffiltext{1}{Bolton Fellow}
\author{Evan D. Skillman}
\affil{Department of Astronomy, University of Minnesota,\\ 116 Church St. 
S.E., Minneapolis, MN 55455}
\email{skillman@astro.umn.edu}
\author{St{\' e}phanie C{\^ o}t{\' e}}
\affil{Canadian Gemini Office, Herzberg Institute of Astrophysics,\\ National 
Research Council of Canada,\\ 5071 West Saanich Road, Victoria, BC, Canada, 
V9E 2E7}
\email{stephanie.cote@nrc.ca}
\begin{abstract}

We present new multi-configuration \HI\ aperture synthesis imaging of the 
nearby dwarf starburst galaxy NGC\,625 obtained with the Australia Telescope 
Compact Array.  Total \HI\ column density images show gas well-aligned with the
optical major axis, and low-column density \HI\ extending to $>$ 6 optical 
scale lengths.  The \HI\ velocity field, on the other hand, is highly 
disturbed, with neutral gas at nearly all detected velocities within the 
central region.  After considering various interpretations, we find that a 
blowout scenario most accurately describes the data. Since at our resolution 
we do not detect any large evacuated holes in the \HI\ disk, we interpret this 
blowout to be the result of the extended (both spatially and temporally) star 
formation event which NGC\,625 has undergone in the last 100 Myr.  This is one 
of the clearest examples of \HI\ outflow detected in a dwarf galaxy.  We find 
no obvious external trigger for this extended star formation event.  We detect 
strong radio continuum emission from the largest \HII\ regions; comparing to 
our HST and ground-based \halpha\ fluxes suggests either appreciable amounts 
of extinction toward the star formation regions, or the contribution of 
non-thermal sources to the radio continuum luminosity.  

\end{abstract}						


\section{Introduction}
\label{S1}

Nearby dwarf galaxies offer the opportunity to study the dynamics of 
galaxy evolution at relatively high spatial resolution.  In these systems we
can probe the nature of star formation and observe its direct impacts on the 
ISM.  Furthermore, the smaller scales of dwarf systems allow us to address the 
global impacts of starbursts on the ISM, and in relatively high-luminosity 
cases, the effects on the surrounding IGM as well.  In these systems we can
attempt to ascertain the triggering mechanisms of starbursts in low-mass
galaxies, which will be an important component of models of the chemical and 
dynamical evolution of galaxies.

As a result of the relative paucity of strong starbursting dwarf galaxies in 
the local volume, there are few detailed studies of the neutral gas kinematics
of such systems.  \citet{kobulnicky95} found the surprising result that in the
well-studied dwarf starburst NGC\,5253, the large-scale rotation of the galaxy 
appears to align along the galaxy's optical major axis.  This rare kinematic 
alignment may be the signature of recent tidal interaction and may provide 
evidence for a triggered starburst in this system.  A more clear case of a 
triggered starburst is found in NGC\,1569.  \citet{stil02c} find that a 
low-mass companion seems to be interacting with the system, and that large 
amounts of the total \HI\ mass are present at unusually high velocities with 
respect to systemic.  These data suggest that infall/outflow and/or tidal 
effects may be important for this system, and that the current starburst may 
present a case of triggered star formation. This burst is having a dramatic 
effect on the ISM, as turbulent motions appear to support the inner disk, and 
the mean velocity dispersion is high ($\sigma_V \sim$ 20\,\kms) compared to 
other starbursts.  

Not all starbursts in low-mass galaxies appear to be tidally induced, however.
In NGC\,1705, well-known because of its spectacular galactic wind, the global 
\HI\ kinematics appear to be quite regular {(Meurer, Staveley-Smith, \& 
Killeen 1998)}\nocite{meurer98}.  While high-velocity \HI\ gas does exist, 
the kinematics suggest simple rotation and a relatively quiescent
velocity field. These few local blue compact dwarfs (BCD's) support the 
conclusions of \citet{taylor97}, wherein it was found that $\sim$\,60\%\ of 
such systems harbor nearby \HI\ companions that may have triggered the 
current starbursts.  However, the local sample also suggests that there must 
exist another mechanism with which to initiate powerful starbursts in 
low-mass galaxies.  In this investigation we will examine the neutral gas 
properties of the nearby dwarf starburst galaxy NGC\,625, with one of the 
primary goals being to discern the nature of the triggering mechanism of 
the current starburst.

The ejection of material from galaxies (i.e., ``blowout'') is a topic of great 
importance for their overall evolution.  In systems of decreasing total mass, 
temporally and spatially concentrated massive star formation
becomes increasingly important.  For the lowest mass galaxies, 
``starbursts'' can dominate the energetics and luminosity of the system
\citep{dekel86,deyoung94}.  Such events can also affect their immediate 
surroundings by injecting energy into the ISM.  If the energies are high 
enough, they can eventually break out of the disk and into the halo, releasing
hot gas and processed material which may or may not escape the system 
entirely to join the surrounding IGM. Numerical simulations of mass loss in 
low-mass systems \citep[e.g.,][]{maclow99} suggest that it is quite 
difficult to remove substantial fractions (i.e., \gsim\ 10\%) of the ISM in 
galaxies with masses $\gsim$ 10$^7$ \msun.  Thus it is useful to examine the 
kinematics of gas in low-mass systems where this model can be empirically 
tested.  As will be discussed below, these observations add NGC\,625 to the 
short list of dwarf systems where neutral gas outflow has been detected.

NGC\,625 is an intriguing starburst system located on the far side of the
nearby Sculptor Group (D = 3.89$\pm$ 0.22 Mpc; {Cannon \etal\ 
2003}\nocite{cannon03}).  In that investigation, HST/WFPC2 photometry was used
to construct dereddened V vs. (V$-$I) color-magnitude diagrams.  It was found 
that NGC\,625 has been undergoing a prolonged ($\sim$ 100 Myr) starburst 
episode.  The spatially resolved nature of those data also allowed us to 
show that the recent star formation has been widespread throughout the 
central regions. Comparing the locations of young stars with the smooth 
distribution of older red giant stars, it is likely that star formation has
been active for an even longer period.  This is one of only a few 
cases where an extended burst of star formation has been observed in a nearby 
dwarf galaxy.  The contributing factors to such an evolutionary scenario are 
not yet fully understood, but the neutral gas dynamics offer us a 
second avenue with which to investigate the recent evolution of this system.

Previous \HI\ observations of NGC\,625 were presented by {C{\^ o}t{\' e}, 
Carignan, \& Freeman (2000)}\nocite{cote00}.  They identified multiple 
emission peaks in the central regions of the galaxy and rotation about the 
galaxy's optical major axis (as opposed to the typical rotation about the 
minor axis).  The complex kinematics precluded the 
derivation of a mass model or a rotation curve for this system and prompted the
present investigation. As the cause for the complex neutral gas kinematics, 
\citet{cote00} suggest a merger scenario - perhaps with a large \HI\ cloud 
\citep[which are numerous in this region of the sky, although mostly at 
differing velocities; see][]{putman02}.  In a study of the ionized gas 
kinematics, however, \citet{marlowe97} also find complex orbits and motions, 
suggestive of a major merger as the cause of the burst.  The possibility does 
exist, interestingly, that the kinematics of the ionized and neutral gas are 
caused by the current burst itself; i.e., perhaps the disk of this system is 
being disrupted by the strong and extended star formation episode.  With these
\HI\ data, we seek to further constrain the nature of the 
recent activity in this system.

Other investigations also point toward the importance of NGC\,625 as a nearby 
example of the dwarf starburst phenomenon.  The spectroscopic investigation of 
{Skillman, C{\^ o}t{\' e}, \& Miller (2003b)}\nocite{skillman03b} finds strong,
broad (5.5\,\AA\ EW) $\lambda$\,4686 \AA\ emission arising from the largest
\HII\ region, suggesting a Wolf$-$Rayet (W$-$R) classification for the current 
burst \citep{conti91} and an age of 4$-$6 Myr.  This result is somewhat 
surprising, given the extended nature of the burst derived in 
\citet{cannon03}, and suggests that the presence of W$-$R features does not 
necessarily imply a young ($<$ 6 Myr) burst. {Skillman, C{\^ o}t{\' e}, \& 
Miller (2003a)}\nocite{skillman03a} find a current star formation rate of 0.05
\msun\,yr$^{-1}$, comparable to the values found in other high-luminosity
dwarf starburst systems.  \citet{bomans98} find diffuse 0.1$-$2.4 keV x-ray 
emission coincident with the large \HII\ region and extended
above the northern side of the plane, suggesting that a galactic wind has been
active or has recently terminated.  The above properties imply that the 
recent star formation has been intense in NGC\,625; the evolution of this 
system has been complex and offers an ideal opportunity to study many 
important questions about the starburst phase of dwarf galaxy evolution.  
We therefore have begun a multiwavelength study of this system, of which these
\HI\ observations form an important part.

For reference, in Table~\ref{t1} we summarize basic observational properties 
of this system.  We organize this paper as follows.  In \S~\ref{S2} we 
discuss the data obtained and our reductions.  The \HI\ data and its analysis, 
our interpretations, and a simple model are discussed in \S~\ref{S3}.  In 
\S~\ref{S4} we discuss the strong radio continuum emission detected from the 
three highest-surface brightness \HII\ regions, and \S~\ref{S5} summarizes 
our conclusions.

\placetable{t1}

\section{Observations and Data Reduction}
\label{S2}

We obtained Australia Telescope Compact Array (ATCA)\footnote{The Australia 
Telescope is funded by the Commonwealth of Australia for operation as a 
National Facility managed by the Commonwealth Scientific and Industrial 
Research Organisation} synthesis imaging, utilizing a number of arrays, 
between 2001 May and 2001 December (see Table~\ref{t2}) for program C968.  
In total, 75.3 hours were spent on source, with 83 independent baselines 
sampling UV distances between 0.139 and 27.1 k$\lambda$.  The data 
were acquired with a correlator configuration that yields two independent 
datasets: broadband continuum data with total bandwidth of 128 MHz, centered 
at 1384 MHz, and narrowband spectral line data with 1024 channels, each with 
an unsmoothed velocity resolution of 0.8 \kms.  For the latter, the central 
observing frequency was 1419 MHz, corresponding to a velocity of 297\,\kms.  
The heliocentric systemic velocity of NGC\,625, 413$\pm$5 \kms, 
corresponds to a 
frequency of 1418.5 MHz, and thus falls near the center of our spectral 
observing frequency band.  Reductions of each of these datasets followed the 
standard format, using the MIRIAD\footnote{See 
http://www.atnf.csiro.au/computing/software/miriad} package; we briefly 
summarize the data handling below.

\placetable{t2}

For the continuum observations, interference and bad data were first removed 
from the UV data.  Observations performed with the sun above the horizon were 
discarded.  Bandpass, flux, gain and phase calibrations were then applied, 
derived from observations of PKS\,1934-638 (primary calibrator) and 
PKS\,0201-440 or PKS\,0104-408 (secondary calibrators). The calibrated UV data 
were then imaged and cleaned to produce the continuum map.  The resulting 
image, produced using natural weighting, is discussed further in \S~\ref{S4}. 
The beam size is 6.8\arcsec$\times$4.8\arcsec, and the rms noise is 
38\ujpbeam. This noise level provides sensitivity to the highest surface 
brightness \HII\ regions in NGC\,625, and we unambiguously detect all three 
of these sources.  

The reduction of the spectral line data is similar to that of the continuum.  
Imaging produces two spectral cubes for analysis, one of high ($\sim$ 
22.5\arcsec) and one of low ($\sim$ 45\arcsec) resolution, both produced using 
tapered uniform weighting, and Hanning smoothed to 2.5 \kms\ velocity 
resolution.  The former will be used to discern the small-scale structure of 
the high-column density neutral gas, while the latter maximizes our 
sensitivity to large-scale structure.  In the former cube, a typical 
line-free single-plane rms noise value is 1.1 \mjpbeam\ (1.55 K), while in the 
latter it is 2.3 \mjpbeam\ (0.85 K).  We discuss the spectral data in the next 
section.

\section{\HI\ Data: Characteristics, Morphology \&\ Kinematics}
\label{S3}

\subsection{\HI\ Characteristics \&\ Morphology}
\label{S3.1}

\begin{figure}
\plotone{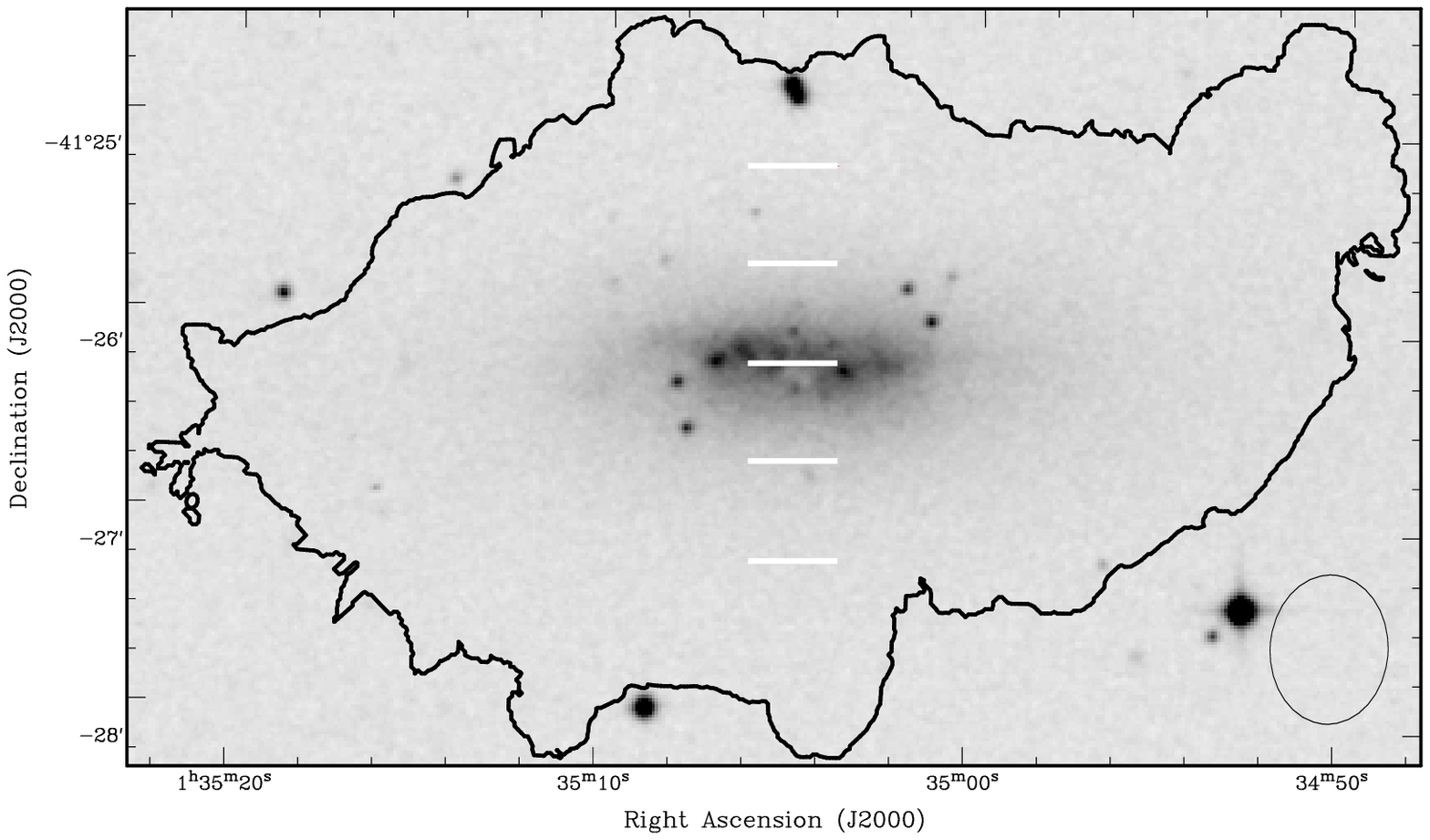}
\caption{POSS\,II near-infrared image of NGC\,625 overlaid with the 
lowest contour of the zeroth-moment (total \HI\ column density) image 
presented in Figure~\ref{figcap2}.  This sensitivity limit corresponds to a 
column density of 1.56$\times$10$^{19}$\,cm$^{-2}$.  The neutral gas is highly
extended in this system; \HI\ is strongly detected to $\sim$ 6 
optical scale lengths \citep{marlowe97}.  The major star formation region is 
evident on the eastern side of the main disk (see also Figure~\ref{figcap6}).}
\label{figcap1}
\end{figure}

\begin{figure}
\epsscale{0.9}
\plotone{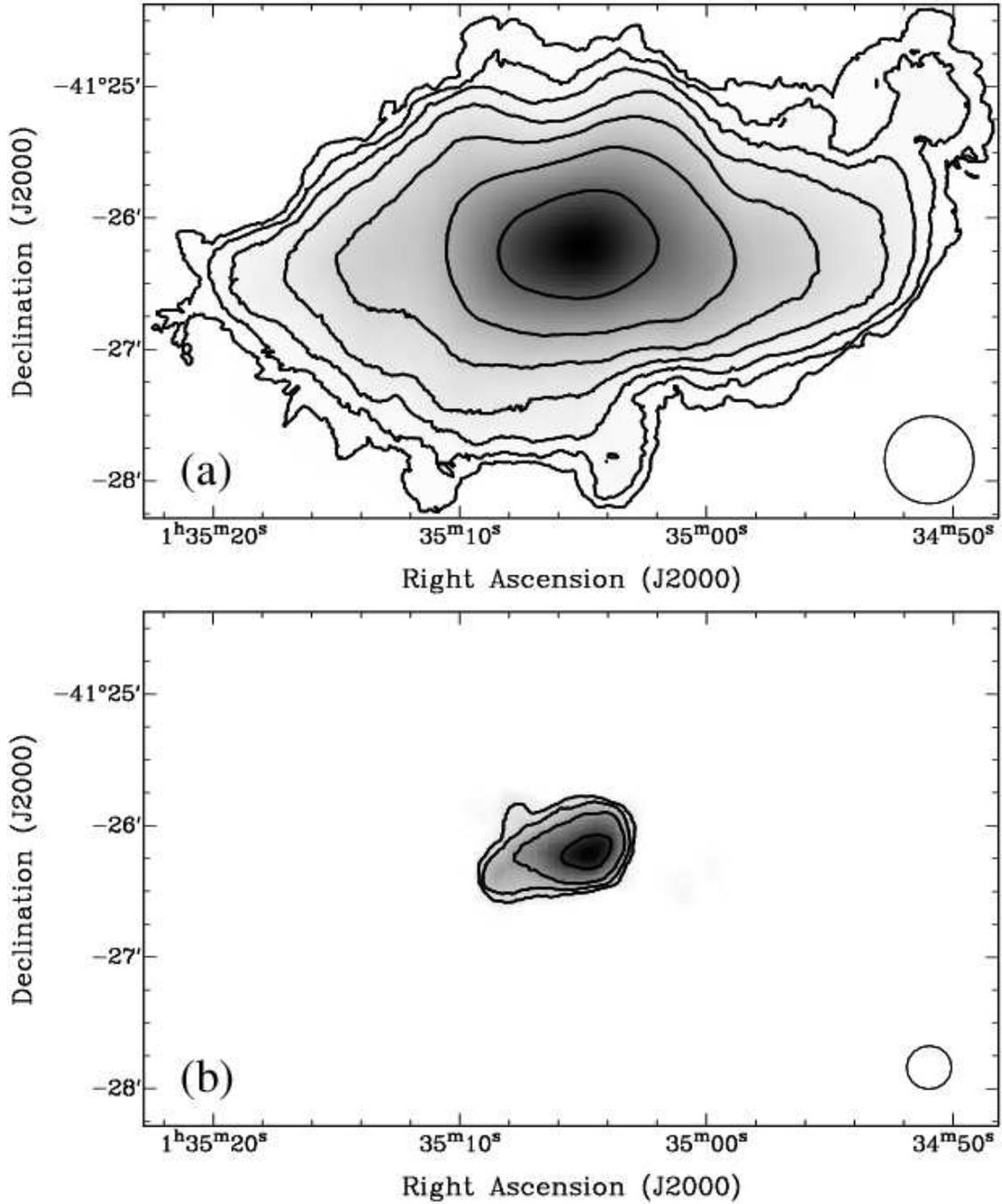}
\caption{Velocity-integrated zeroth-moment images, representing total \HI\ 
column density.  (a) shows the low-resolution (45\arcsec\ beam) image, while
(b) shows the higher-resolution (22.5 \arcsec\ beam) image.  Contours in (a) 
show levels of 1, 2, 4, 8, 16, 32 and 64\%\ of the peak intensity 
(1178 K$\cdot$\kms, or 2.15$\times$10$^{21}$ cm$^{-2}$), while the 
contours in (b) show levels of 10, 20, 40 and 80\%\ of the peak intensity 
(1321 K$\cdot$\kms, or 2.41$\times$10$^{21}$ cm$^{-2}$).  The beam size
and shape is shown at the bottom right of each image.}
\label{figcap2}
\end{figure}

In Figures~\ref{figcap1} and \ref{figcap2} we present the total \HI\ emission 
as derived from zeroth moment calculations from the spectral line data cubes.  
Figure~\ref{figcap1} displays a comparison between the optical galaxy and the
extent of the \HI\ at the sensitivity limit of our observations. At the 
resolution available, the \HI\ in NGC\,625 aligns well with the optical galaxy 
with a low column density extension toward the northwest \citep[also seen 
in][]{cote00}.  These observations detect \HI\ to more than 6 optical scale 
lengths \citep[33\arcsec;][]{marlowe97}.  While this is smaller 
than the extent of the large \HI\ halos seen in some dwarf galaxies (e.g., 
NGC\,4449, {Bajaja, Huchtmeier, \& Klein 1994}\nocite{bajaja94}; {Hunter 
\etal\ 1998}\nocite{hunter98b}), the total \HI\ envelope is comparable in 
size to those of the dwarf and low-surface brightness galaxies in the 
samples of \citet{vanzee97b} and \citet{cote00}.

Because the \HI\ distribution ellipticity and orientation agree with those of 
the optical distribution, one expects a velocity field typical for a dwarf 
galaxy (i.e., nearly solid-body rotation about the optical minor axis).  
However, as demonstrated by \citet{cote00}, the velocity field for NGC\,625 
is anything but typical.  We discuss the velocity structure in detail in 
\S~\ref{S3.2}.

Figure~\ref{figcap2} presents zeroth-moment images at both low and high 
resolution.  Both zeroth moment images show a peak column density $>$ 
2$\times$10$^{21}$ cm$^{-2}$.  The main \HI\ peak is not coincident with the 
location of the current major star formation region, but rather is offset to 
the east by $\sim$ 300 pc.  It is interesting to note that we detect a sizable 
high-column density (N$_H$ $>$ 10$^{21}$ cm$^{-2}$) region in the disk.  This 
suggests plentiful dense gas with which to sustain an extended burst of star 
formation, as was derived in \citet{cannon03}.  

From the low-resolution data cube, we derive a total \HI\ gas mass of 
(1.1$\pm$0.2)$\times$10$^8$ \msun. We note, however, that there exists 
significant (maximum A$_V$ $=$ 0.57 mag.) and variable (differences of $\sim$ 
0.2 mag. over physical scales of tens of parsecs) extinction in the disk 
\citep{cannon03}, and likely a sizable molecular gas reservoir (C{\^ o}t{\' 
e}, Cannon, \& Braine, in preparation), suggesting that the total gas mass 
is considerably higher than this value.  This mass measurement is in good 
agreement with that found in \citet{cote00} after correcting for the new 
distance found in \citet{cannon03}.

\subsection{The Disturbed \HI\ Kinematics}
\label{S3.2}

\begin{figure}
\plotone{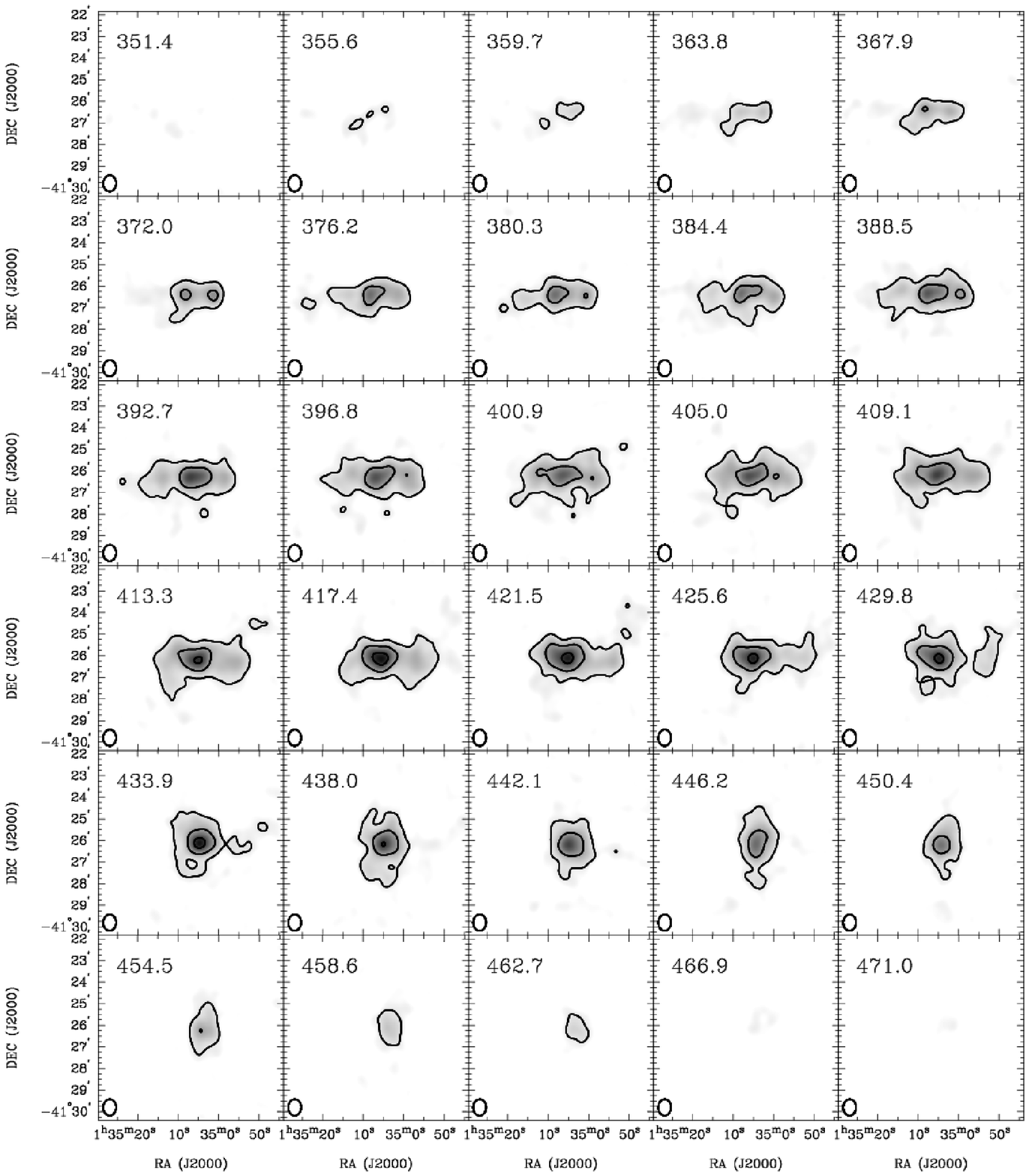}
\caption{Low-resolution (45\arcsec\ beam) \HI\ data showing planes of 
constant velocity, as labeled in the upper left corner of each frame.
The image stretch is linear, from the noise level (2.3 \mjpbeam\ 
in line-free individual planes) to the maximum intensity of 53 
\mjpbeam.  The contours show levels of 3, 9, and 18\,sigma, corresponding to
1.18, 3.54, and 7.08$\times$10$^{19}$ cm$^{-2}$, respectively.  The figure 
shows three-channel averages, separated by 5 increments of the unsmoothed 
velocity resolution of 0.8 \kms.}
\label{figcap3}
\end{figure}

We next examine the velocity structure of the \HI\ gas in NGC\,625.
In Figure~\ref{figcap3} we present individual channel maps from the
low-resolution cube.  At multiple velocities, the \HI\ gas is
multi-peaked, and in some cases (most notably 429.8 \kms) even forms
spatially distinct structures at this sensitivity level.  Indeed, as
will be argued below, this system appears much more complex in
velocity space than the initial distribution and column density images
(Figures~\ref{figcap1} \&\ \ref{figcap2}) would suggest.

A closer examination of the channel maps in Figure~\ref{figcap3} shows
that NGC\,625 is undergoing solid-body rotation about its optical
minor axis (i.e., as expected for an inclined rotating disk).
However, superposed on this disk are a complex variety of kinematic
structures that make the interpretation of the velocity field
difficult.  Some of this structure (especially at high velocities with
respect to systemic) makes the first moment images (see below)
appear to contain a component of rotation about the galaxy's optical
minor axis.  This unusual behavior, initially posited by
\citet{cote00}, appears to be the result of the superposition of
multiple velocity components along the line of sight and not the
result of actual rotation about the optical minor axis.

\begin{figure}
\plotone{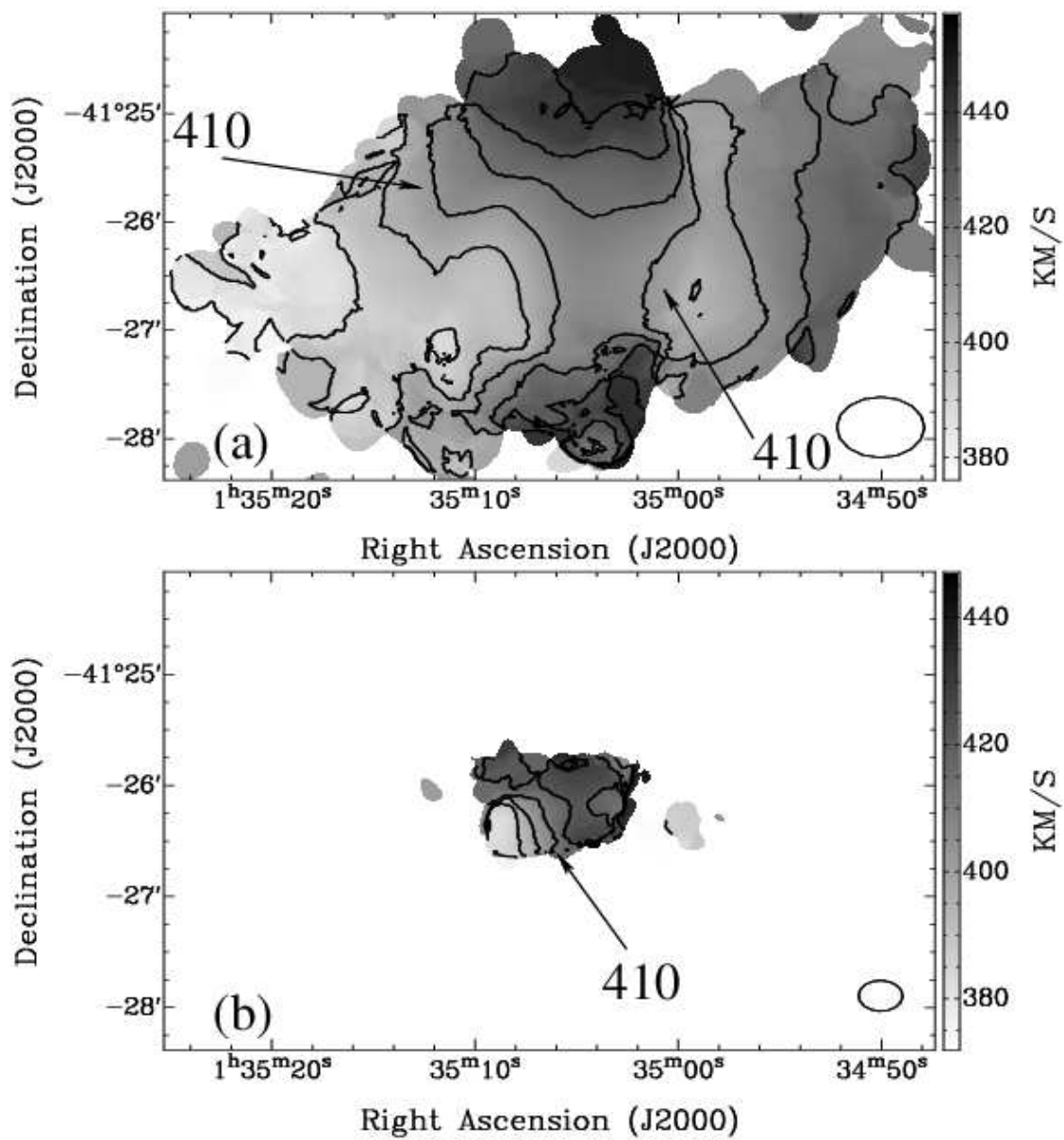}
\caption{Velocity-integrated first-moment images, representing radial velocity.
(a) shows the low-resolution (45\arcsec\ beam) velocity field, while
(b) shows the higher-resolution (22.5 \arcsec\ beam) velocity field.  The 
contours are spaced by 10 \kms\ intervals, ranging from 380 \kms\ to 440
\kms\ in (a) and from 380 \kms\ to 430 \kms\ in (b).  The velocity field is 
highly disturbed, with components across the full velocity range within the 
disk, as well as components with steep velocity gradients.}
\label{figcap4}
\end{figure}

The first moment images (representing intensity-weighted velocity field) are
shown in Figure~\ref{figcap4}.  In the low-resolution image, it is
clear that many velocity components have been averaged together along
the line of sight to form the column density
images shown previously.  Note in particular the high-velocity gas
seen both above and below the \HI\ major axis and the lower-velocity
gas seen on both sides of the optical center.  The high-resolution
velocity field only provides information on the highest-column density
\HI\ gas, but there does appear to be agreement in the velocity
structure in the regions detected in both the low- and high-resolution
velocity fields.  Furthermore, solid-body rotation appears to be evident 
in both cubes, and the high-resolution velocity field shows the central 
regions less confused with the high-velocity gas above and 
below the disk as seen in the lower-resolution cube. 
This very confused velocity structure warrants a further
investigation of the nature of the \HI\ emission at all velocities.

\begin{figure}\vspace{-3 cm}
\epsscale{0.8}
\plotone{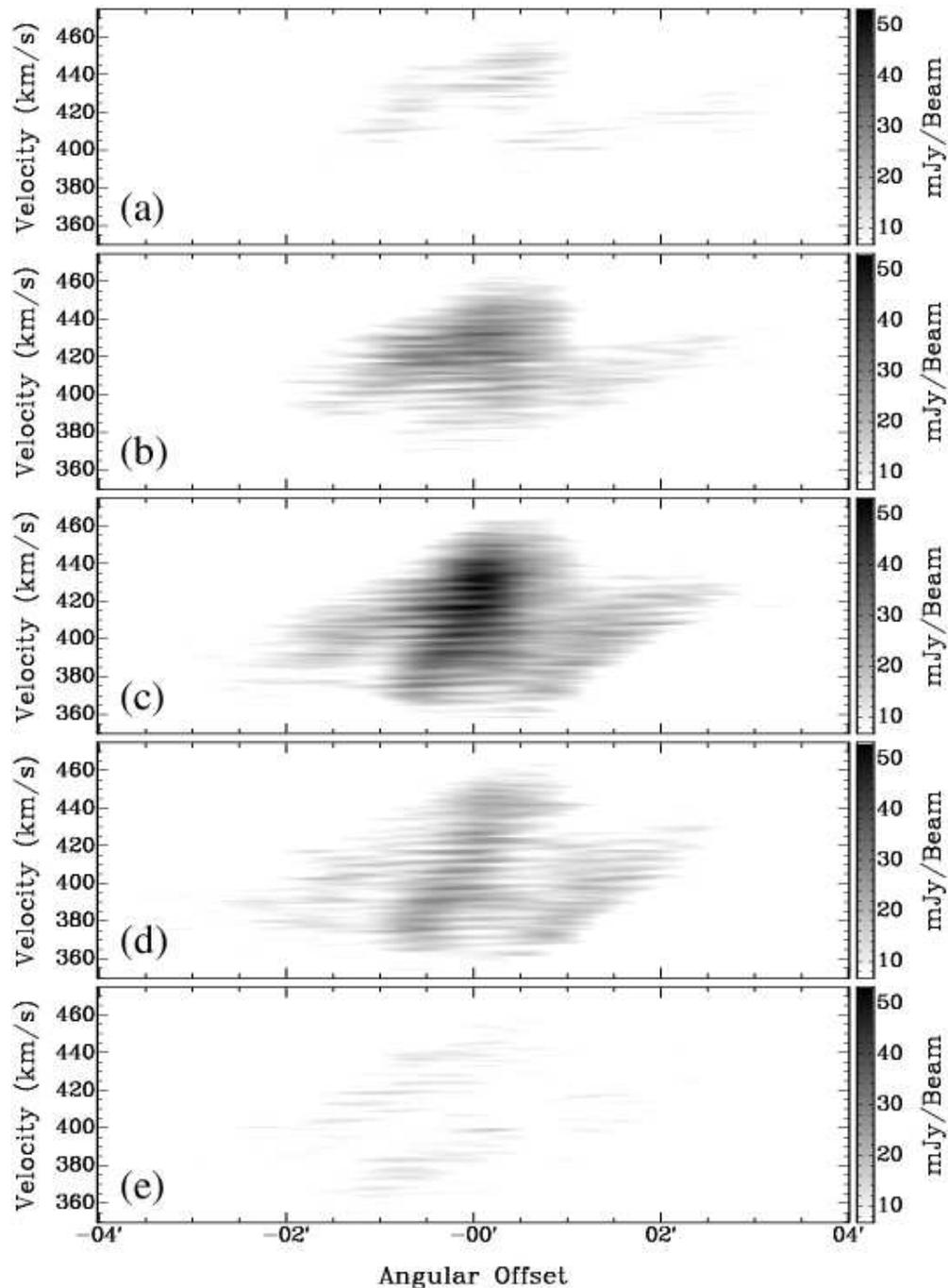}\vspace{-0.5 cm}
\caption{Five independent major-axis PV diagrams, taken at different distances
from the \HI\ column density maximum, located at $\alpha$,$\delta$ (J2000) $=$
01:35:4.93, $-$41:26:12.20.  (a) is located 60\arcsec\ N of the disk, (b) at
30\arcsec\ N, (c) coincident with the \HI\ peak, (d) at 30\arcsec\ S, and (e)
at 60\arcsec\ S.  Each of these locations is shown by a small white line in
Figure~\ref{figcap1}.  The cuts are 8\arcmin\ long, at a position angle of 
90\degree.  Near the disk, the 
kinematic signature of blowout dominates the \HI\ velocity structure.}
\label{figcap5}
\end{figure}

In Figure~\ref{figcap5} we present five position-velocity cuts along
the direction of the optical major axis (position angle $=$ 90\degree,
width $=$ 3\arcsec).  One is centered on the \HI\ column density peak,
two at $\pm$ 30\arcsec, and two at $\pm$ 60\arcsec\ north and south of
this position (note in Figure~\ref{figcap1} the five white lines,
which denote the center of each of these PV cuts).  Each PV cut is
separated by nearly a full beam width.  In each section of the figure,
note the clear multi-component emission seen at different positions
along the cut.

Beginning with the cut through the \HI\ column density maximum
(Figure~\ref{figcap5}c), note the prominent detection of gas at nearly
the full range of velocities (360 $-$ 460 \kms) which is coincident
with the main body of the galaxy (i.e., within 1\arcmin\ of the
central location, and within the optical extent of the system).  There
is an identifiable, very steep velocity gradient (increasing
velocities from negative to positive offsets), which on first
inspection might be interpreted as the signature of solid-body
rotation of at least part of the \HI\ disk.  However, two lines of
evidence argue against this.  First, under the assumption that this
disk is rotationally supported, the implied central mass is $>$
4$\times$10$^8$ \msun; i.e., much larger than the kinematic mass of
the system (see further discussion in \S\S~\ref{S3.3.1},
\ref{S3.3.3}).  Second, and more importantly, the fact that there
exists \HI\ at such a wide range of velocities in the central region
argues for a highly turbulent disk; this may be a result of the
extended and intense star formation event which this system has
undergone, depositing energy throughout the disk without the creation
of a single massive evacuated cavity as a result.  We make further
projections on the nature of this gas in the following section.  The
confused nature of the \HI\ gas in the central regions prevents us
from deriving an unambiguous rotation curve for this system.

Next, note the clear detection of two kinematically distinct features
in all five PV diagrams, reaching $>$ $\pm$ 2\arcmin\ from the central
position.  There is a clear velocity discontinuity at an angular
offset of $-$1\arcmin, where a change of nearly 100 \kms\ occurs in a
very short distance.  Similarly, a velocity discontinuity is apparent
at an angular offset of $+$ 1\arcmin.  The symmetry of these features
suggests that these are parts of the regularly-rotating \HI\ disk of
NGC\,625.  The high and low-velocity gas both north and south of the
central component appears to be superposed on this apparently
solid-body rotating component.

The unprecedented structure evident in these PV cuts is a kinematic
signature consistent with active blowout of neutral gas from the disk
of NGC\,625.  The presence of \HI\ gas at all velocities in the
central sections of the galaxy suggests that large amounts of energy
have been deposited into the ISM of this system.  Some of the most
conspicuous structure is present in the velocity region 430 $-$ 450
\kms, as seen in the panels of Figure~\ref{figcap3}.  This
high-velocity gas appears both above and below the optical extent of
the system and provides strong evidence for outflow behavior (see
\S~\ref{S3.3} for further discussion).

While the strongest tracer of blowout is usually \halpha\ kinematics,
this study adds NGC\,625 to the short list of systems where \HI\
blowout is unambiguously detected (see also the {Ott \etal\
2001}\nocite{ott01} study of Ho-I).  Certainly violent star formation
has played a role in the evolution of similar systems [e.g., the SMC,
{Stanimirovic \etal\ 1999}\nocite{stanimirovic99}; Ho-II, {Puche
\etal\ 1992}\nocite{puche92} (but see also {Bureau \& Carignan
2002}\nocite{bureau02}); NGC\,1569, {Stil \& Israel
2002}\nocite{stil02c}; NGC\,1705, {Meurer \etal\
1998}\nocite{meurer98}].  While we do not see the small-scale effects
of active star formation on the ISM as clearly in NGC\,625 as in these
other systems (e.g., no clear cases of evacuated shells or chimneys),
it is clear that the current active star formation has had a dramatic
effect on the ISM.  As we argue in the following section, we interpret
this kinematic signature as active blowout of large amounts of \HI\
gas from the disk of NGC\,625.  This blowout may be the result of the
spatially and temporally extended star formation episode which
NGC\,625 has undergone in the last 100 Myr.  Higher-resolution \HI\
imaging would likely show that the active star formation is in the
process of evacuating small areas of the ISM.  In the next section we
consider various interpretations of these PV diagrams. 

\subsection{Possible Kinematic Models for NGC\,625}
\label{S3.3}

Here we consider a large range of possible interpretations of the disturbed 
kinematics in NGC\,625. In the end, we find that blowout superposed against a 
disk in nearly solid-body rotation is the only model which successfully 
reproduces all of the characteristics of the observations.  While these data 
are not of sufficient resolution to detect the small-scale (e.g., $<$ 100 pc) 
features of the ISM (where many \HI\ holes and other features may be found), 
they are more than adequate to examine the nature of the large-scale \HI\ 
features which are associated with this putative blowout model.

\subsubsection{A Simple Rotating Disk}
\label{S3.3.1}

As noted above, the general kinematics of the \HI\ in NGC\,625 are consistent
with a slowly rotating dwarf galaxy, observed at a relatively high inclination.
As seen in Figure~\ref{figcap3}, the channel maps are consistent with an \HI\ 
disk that retains strong components of rotation about the galaxy's optical 
minor axis.  However, superposed on this simple disk model are complicated 
velocity components that indicate a highly irregular integrated velocity field,
and neutral gas at nearly all detected velocities within the \HI\ disk.  As 
can be seen in Figure~\ref{figcap5}, the simple solid-body behavior is 
prominent from $-$2\arcmin\ to $+$2\arcmin;  however, in the inner $\pm$ 
1\arcmin, the superposition of velocity components interrupts this simple model
of the \HI\ distribution.  Therefore, these observations are not consistent 
with {\it only} a simple \HI\ disk in solid-body rotation.  Rather, a more 
complicated model is required to explain the observations. 

\subsubsection{Blowout Plus Rotation}
\label{S3.3.2}

The disk of NGC\,625 is seen nearly edge-on.  From our HST/WFPC2 observations,
the optical axis shows an inclination of 69\degree\ \citep{cannon03}.  
Since we cannot unambiguously derive a rotation curve, we cannot derive a 
precise value for the inclination of the galaxy using these \HI\ data alone.
However, using elliptical fits to \HI\ column density contours suggests an
inclination of (65$\pm$5)\degree.  For the purposes of this simple model, we
adopt the value of 65\degree\ as the inclination of the disk.  

Let us assume a ``standard'' conical blowout emanating from the center of the
\HI\ disk ({Mac~Low, McCray, \& Norman 1989}\nocite{maclow89}; {Mac~Low \& 
Ferrara 1999}\nocite{maclow99}).  This blowout will push neutral gas 
preferentially in the direction of lowest pressure; i.e., directly out of the
disk (N/S in the case of NGC\,625).  Assuming this gas breaks out of the \HI\ 
disk (which should be in solid-body rotation), it will then allow hot gas and 
metals to be vented into the galactic halo.  

In a major-axis position-velocity slice, this scenario will appear similar 
to Figure~\ref{figcap5}.  More explicitly, the expected solid-body rotation of 
the \HI\ disk will appear as a diagonal feature, from low velocities and large
separation on one side of the disk, to high velocities and large separation on
the opposite side of the disk.  Superposed on this will be gas of all 
velocities near the location of the galactic wind center; i.e., at the location
of the starburst, we expect to see gas of all velocities, since in a conical 
blowout model, there will exist some component of velocity along our line of 
sight from all 360\degree\ of the expanding material.

By observing position-velocity diagrams at varying distances from the central
disk (here, at different declinations), we should see the solid-body component
decrease in strength, while the gas above and below the disk should remain 
prominent.  This is caused by two factors.  First, the disk itself 
is of finite thickness.  Thus, the further the distance from the main \HI\ 
disk, the less neutral gas is involved with the total system rotation.  Second,
the gas which is being pushed out of the disk should remain visible
until the scale height of the blowout.  Once this position is reached, the 
intensity in these features should quickly decrease.

Each of the main points of the blowout scenario is observed in 
Figure~\ref{figcap5}.  Furthermore, we note that our measure of the 
inclination is consistent with more \HI\ above the disk than below (compare 
high and low velocities in the central region cut, and high and low velocities
between the $+$ 60\arcsec\ and $-$ 60\arcsec\ cuts).  Thus this series of PV 
diagrams appears to be consistent with a simple picture of a normal dwarf 
galaxy \HI\ disk, highly inclined with respect to the line of sight, with an 
internal wind that is causing some of the \HI\ gas to be vented both above and
below the plane (with more prominent emission above than below the disk).

\subsubsection{Superposed or Counter-Rotating Disks}
\label{S3.3.3}

Superposed disks (a kinematically distinct small disk rotating in the same 
sense as a larger disk) have the potential to reproduce the main features of 
the PV diagram.  Here, a large, rotating halo will produce the \HI\ gas 
detected between $-$2\arcmin\ and $+$2\arcmin\ in the PV diagram.  Superposed
on this will be a second component, rotating in the same sense and thus 
appearing in a similar orientation in the PV diagram.  Depending on the 
difference in mass and rotation speed, the position angle of these two 
components will vary in the PV diagram.  Thus, this model appears to be able
to describe the observed PV diagrams.

Upon a closer inspection of Figure~\ref{figcap5}, however, we can eliminate 
this possibility because of the size of the components needed.  In order to 
reproduce the observed PV diagram the enclosed mass in the smaller central 
component would have to be $>$ 4$\times$ 10$^8$ \msun. This greatly exceeds 
the masses within the inner 1 kpc found consistently in dwarfs and irregulars, 
based on mass models of their rotation curves that include gas, stellar and 
dark matter contributions \citep[e.g.,][]{cote95}.

The observed PV diagrams are also inconsistent with a counter-rotating disk
(e.g., as seen in NGC\,4449, {Bajaja \etal\ 1994}\nocite{bajaja94}; {Hunter 
\etal\ 1998}\nocite{hunter98b}).  There, the orientation of the two components 
(disk and halo) must be in the opposite sense in a major-axis PV diagram;  
i.e., the direction of increasing velocity for one must be the direction of 
decreasing velocity for the other.  This is opposite to what is observed.  
Only disks rotating in the same direction can produce the arrangement of 
features seen in our PV diagram; however, as discussed above, this can only 
occur with unphysically large central mass concentrations in this low-mass 
system. 

\subsubsection{Infall}
\label{S3.3.4}

In most galaxies it is impossible to distinguish between infall and outflow.  
If NGC\,625 is currently accreting material from a diffuse halo or neighboring
\HI\ cloud, this gas could appear as the high-velocity gas near the main \HI\ 
disk.  However, the symmetry of the NGC\,625 system is better explained in the 
context of outflow.  If infall were taking place, we would expect to see more 
gas on one side of a system than the other (e.g., as would result from a 
merger or from the accretion of a gas cloud).  The high-velocity gas (with 
respect to systemic) seen both above and below the central regions of NGC\,625
is more likely exiting the system; because symmetric infall is very rare, this 
scenario will have difficulty in explaining the observed PV diagrams and 
velocity field.

\subsubsection{``Ongoing Assimilation''}
\label{S3.3.5}

Some nearby dwarf irregular galaxies are known to have very extended \HI\ 
structures with complex kinematics (e.g., IC\,10, {Shostak \& Skillman 
1989}\nocite{shostak89}, {Wilcots \& Miller 1998}\nocite{wilcots98}; 
NGC\,4449, {Hunter \etal\ 1998}\nocite{hunter98b}).  \citet{hunter98b}
invented the phrase ``ongoing assimilation'' to present a scenario which best 
describes their present status.  In that investigation, large 
``streamers'' of \HI\ gas, highly structured and extending away from the main
disk of the galaxy, were posited to be primordial inhomogeneities in the gas 
cloud surrounding the system that are currently being incorporated into the 
galaxy.  Such a delayed formation scenario had been suggested theoretically
{(Silk, Wyse, \& Shields 1987)}\nocite{silk87}, and the presence of 
considerable amounts of diffuse gas in the NGC\,4449 halo lends credence to 
this scenario.  

Such a scenario presents an alternative interpretation of these \HI\ 
observations of NGC\,625.  Evidence for ongoing assimilation comes from the 
large amount of diffuse \HI\ surrounding the main galaxy.  However, this 
scenario is not able to explain the complicated velocity structure seen in 
Figure~\ref{figcap5} for two reasons.  First, there is no coherent spatial 
structure in the diffuse \HI\ gas surrounding the galaxy, as seen in NGC\,4449.
This implies that there has been no ``trigger'' for the assimilation of the 
material (which should produce asymmetric structure on large scales), and 
that we have caught this system in a rare case of activity.  Second, and more
importantly, the high-velocity gas is seen comparatively close to the main 
disk of the galaxy, suggesting its energy source is nearby.  Unlike NGC\,4449,
where structures remain coherent in velocity for $\sim$ 200 \kms\ and spatially
over scales of $\sim$ 100 kpc, the \HI\ in NGC\,625 is confined to within 
$\sim$ 50 \kms\ of systemic and to within $\sim$ 2.5 kpc of the optical center
of the system.  Taken together, these points argue that such a scenario likely 
cannot explain the structures seen in Figure~\ref{figcap5}.

\subsubsection{Superposed Anomalous-Velocity Clouds}
\label{S3.3.6}

A chance superposition of two anomalous-velocity clouds could potentially 
explain the complicated PV diagrams.  Indeed, there exist many \HI\ clouds 
coincident with the Sculptor Group; however, they display a wide velocity 
dispersion about the systemic velocity of the group, and none are coincident 
in both position and velocity with NGC\,625 \citep[see][]{putman02}.  While 
there is a chance that two small clouds have escaped detection in the 
compilation of \citet{putman02}, it appears that a chance superposition of 
two major \HI\ clouds is an unlikely scenario with which to explain the PV 
diagrams.

\subsubsection{ISM Turbulence}
\label{S3.3.7}

The observed PV diagrams are inconsistent with those expected from a highly 
turbulent ISM.  While there certainly are turbulent processes at work in any 
area of massive star formation and evolution, it is highly unlikely that 
such random processes could produce the symmetrical distribution of 
high-velocity gas near the center of the galaxy.  If turbulence 
were the cause of the high-velocity gas, it would need to remain coherent over
spatial scales exceeding a kiloparsec (i.e., the separation between the central
PV cut and the cuts at $\pm$ 60\arcsec).  Thus, while turbulence likely has a 
hand in disturbing the small-scale structure of the ISM in NGC\,625, it is 
highly unlikely that it has produced the high-velocity gas north and south of 
the disk as seen in the PV diagrams.

\subsection{Simple Calculation of Burst Energetics}
\label{S3.4}

We interpret the complicated velocity structure in NGC\,625 as the 
signature of \HI\ blowout from the disk.  However, at our resolution, we do 
not detect evacuated holes or shells in the disk that would support such a 
scenario.  This implies either that we are witnessing this blowout event 
relatively early in its evolution, or that the blowout is the result of a 
continuous, widespread star formation event that has deposited 
sufficient energy into the ISM to cause \HI\ gas to be expelled into the halo.
Given our recent star formation history analysis \citep[see][]{cannon03}, we
suggest the second scenario is more likely.  

Based on the results presented in \citet{cannon03}, we can estimate the 
mechanical energy input into the ISM from the recent extended star formation
event. In that investigation the authors found a high, but declining, star 
formation rate over at least the last 100 Myr.  More specifically, the average 
star formation rates over the coarse age bins 0-25 Myr, 25-50 Myr, and 50-100 
Myr were found to be 0.008, 0.01 and 0.04 \msun$\cdot$yr$^{-1}$, respectively.
Using these star formation rates, and assuming an IMF, we can estimate the 
number of SNe that have exploded thus far in this extended star formation 
event.  In particular, we assume a Salpeter IMF, and that all stars above 8 
\msun\ explode as SNe, each depositing an average of 10$^{51}$ erg of 
mechanical energy into the surrounding ISM \citep{burrows00}.  

The results of this simple model suggest that the recent star formation in 
NGC\,625 has deposited $\sim$ 10$^{55}$ erg of energy in the last 100 Myr. If 
the total mass of NGC\,625 is $\sim$ 10$^9$ \msun, and a scale height of 
$\sim$ 1 kpc is assumed (roughly the separation between the major \HII\ region
and the extent of diffuse x-ray emission; {Bomans \& Grant 
1998}\nocite{bomans98}), then this amount of energy is comparable to the total 
gravitational binding energy of the system.  Thus the fact that we see  
kinematic evidence for blowout should not be surprising.  

A calculation of the mass of outflowing material is not straightforward. That 
is, there is no clear kinematic separation between outflowing components and 
stationary gas within the disk.  Rather, it appears that we are viewing a 
superposition of gas at many velocities along the same line of sight, making 
it difficult to accurately disentangle gas that is outflowing from gas that is
stationary within the disk.  However, simple calculations suggest that there 
has been sufficient energy deposited into the ISM to accelerate a substantial
fraction of the neutral gas in this system, with sensible assumptions about
the efficiency of mechanical energy deposition.  For example, assuming a 20\%
efficiency for mechanical energy conversion from input SNe energy and a total
baryonic mass of $\sim$ 10$^9$ \msun\ (although note that the dark matter 
content remains unconstrained from these observations), $\sim$ 20\% of the 
\HI\ mass (or 2.2$\times$10$^{7}$\,\msun) could be moved roughly 1 kpc above 
the disk.  Note that a larger dark matter content will decrease the fraction of
the \HI\ mass that could be accelerated.  Regardless of the mass of material 
that is currently outflowing, the energetics are more than adequate to account
for an outflow scenario.

\subsection{A Simple Case of Blowout?}
\label{S3.5}

As discussed above, we interpret the complex velocity structure in 
NGC\,625 as the signature of blowout, superposed on a normal disk that is 
undergoing solid-body rotation.  This simple scenario offers the best 
description of the observed PV diagrams.  Furthermore, it has been argued that 
other interpretations of the data have shortcomings that preclude them from 
accurately reproducing the PV structures.

We note that this interpretation is consistent with other observations of 
NGC\,625.  First, using the luminous blue helium burning stars, in 
\citet{cannon03} we derived a recent star formation history where sustained 
high star formation rates have been maintained for at least the last 100 Myr.
While spectroscopic W$-$R features suggest a much shorter duration for the 
current burst, the high \HI\ column density shown throughout much of the 
central disk, and the movement of star formation throughout the disk found in
the HST data, suggest that the star formation in this system has been
widespread throughout the disk, rather than confined to the location of the 
current major burst.  The 100 Myr timescale would allow large amounts of 
energy to be put into the ISM, without necessarily leaving one large evacuated 
cavity which would be detectable at our current resolution.  Regardless, we 
know that there has been an extended period of star formation in NGC\,625.

Second, \citet{bomans98} find diffuse soft x-ray emission above the northern 
side of the disk.  This gas shows two x-ray peaks, separated by more than 
1\arcmin.  Based on the pointing accuracies of ROSAT\footnote{See the online
ROSAT Point Sources Catalogue, http://wgacat.gsfc.nasa.gov/wgacat/wgacat.html},
we can be sure that this x-ray emission is located in the northeastern halo 
of the galaxy, and potentially coincident with the disk as well.  
Given the inclination and the orientation of the outflow which we infer from 
the PV diagrams, such hot gas (T $\sim$ 10$^6$ K) would be preferentially 
seen on the northern side of the disk, i.e., exactly where the diffuse gas is 
detected.  For this hot gas to reach the large distances above the disk where 
it is seen, an active outflow must have cleared the way for the hot gas to 
escape.  It is possible that the two x-ray peaks correspond to two separate 
outflow occurrences.

Finally, \citet{skillman03a} find chimney-like \halpha\ filaments extending 
radially upward from the galaxy on the northern side (see regions 11 \& 12
of their Figure\,1).  This is exactly what is expected for a blowout scenario, 
where such chimneys trace the path of hot gas outflow into the halo of the 
galaxy \citep{heiles93}. These low-surface brightness features are clear 
indicators that a low-density region must have been evacuated to allow 
unattenuated \halpha\ emission to escape to large distances above the disk.
Each of these points supports our interpretation of the unusual velocity 
structure as the signature of blowout from the disk of NGC\,625.  

Comparing our observational results with the numerical simulations of mass 
loss in dwarf galaxies in \citet{maclow99}, we find good agreement.  Our \HI\
mass corresponds most closely to the 10$^8$ \msun\ model presented there.  
In \S~\ref{S3.4} we estimated the mechanical energy input into the ISM by the 
recent star formation.  The results suggest a scenario comparable to the 
L$_{38}$ $=$ 1-10 models presented in \citet{maclow99}.  Examining the gas 
distributions presented therein, we find that \HI\ is to be expected tens of 
kpc above and below the disk of the system.  Since our \HI\ distribution is 
smaller than this, we posit that the spatially widespread star formation 
derived in \citet{cannon03} has resulted in outflow, but has deposited this 
energy throughout the disk, rather than in one spatially concentrated area. 
Further agreement is found with the x-ray and \halpha\ emissivities presented
in \citet{maclow99}; the diffuse x-ray and \halpha\ gas discussed above is
consistent with the results for this gas mass and energy input.  

\section{The Radio Continuum Emission and Extinction Toward the Starburst}
\label{S4}

Radio continuum emission can be a useful probe of the physical conditions in
active star formation regions.  In particular, the contribution of non-thermal
sources can be identified based on the spectral index of the emission.  Via
comparison with recombination line intensities (e.g., \halpha, \hbeta),
one can attain a detailed measure of the reddening toward heavily-obscured 
star formation regions that would otherwise be inaccessible at optical 
wavelengths.  

\begin{figure}
\epsscale{1}
\plotone{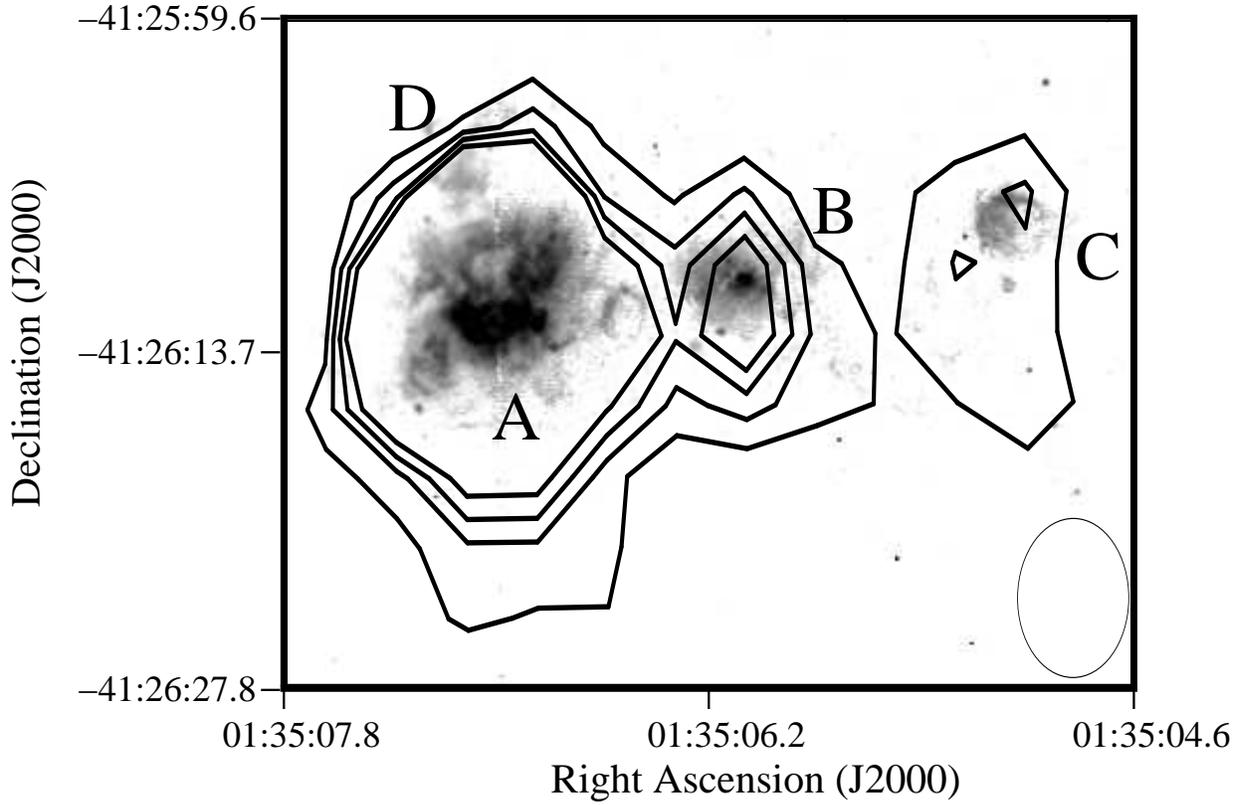}
\caption{1384\,MHz continuum contours overlaid on a continuum-subtracted 
\halpha\ image of NGC\,625 (figure drawn from Cannon \etal\ 2003). Contours 
are at levels of 0.2, 0.3, 0.4, 0.5 mJy/Beam, corresponding to 5, 7.5, 10 
and 12.5\,$\sigma$.  Labels A-D denote the nomenclature from Cannon \etal\
(2003); these complexes can be seen near the eastern end of the disk in 
Figure~\ref{figcap1}.  The beam size is 4.8\arcsec\,$\times$\,6.8\arcsec\ and 
is shown at the bottom right.  The continuum luminosities of these \HII\ 
regions are used in \S~\ref{S4} to estimate the reddening and contribution 
from non-thermal sources to the continuum luminosity.}
\notetoeditor{The Cannon \etal\ (2003) reference in this Figure caption should 
be referenced to Cannon \etal\ (2003).  Using the ``nocite'' command in the 
figure caption causes errors upon compiling the latex file.}
\label{figcap6}
\end{figure}

The ATCA correlator allows concurrent \HI\ and continuum studies of NGC\,625.
The broadband continuum data, centered at 1384\,MHz, allow a detailed look 
at the 20\,cm radio continuum emission arising from the large \HII\ regions 
in NGC\,625.  As shown in Figure~\ref{figcap6}, we detect strong radio 
continuum from the largest \HII\ regions (NGC\,625\,A, B, C of {Cannon \etal\ 
2003}\nocite{cannon03}).  The spectroscopic W$-$R features 
\citep{skillman03b} arise from the strongest radio continuum source, which 
is coincident with \HII\ region NGC\,625\,A.  

The total 1384\,MHz flux densities from the major \HII\ regions (within 
apertures the size of, or larger than, the beam) are given in Table~\ref{t3}. 
Note that \HII\ regions A and D are measured together; since the contribution 
of D to the combined flux is $<$ 4\%, this will not affect the
results presented here.  Following \citet{caplan86}, for a thermal continuum 
source in the presence of no extinction, the \halpha\ flux and the continuum 
flux at 1384 MHz are related by:

\begin{equation}
\frac{j_{H\alpha}}{j_{1384}} = 
\frac{(8.658\times10^{-9})(\frac{T}{10^4})^{-0.44}}{(10.486 + 
1.5\cdot ln(\frac{T}{10^4}))}
\end{equation}

\noindent where j$_{H\alpha}$ is the flux at \halpha\ in 
erg\,sec$^{-1}$\,cm$^{-2}$, j$_{1384}$ is the flux density at 1384 MHz in Jy, 
and T is the electron temperature in K. From \citet{skillman03b} we adopt an 
electron temperature for each \HII\ region, listed in Table~\ref{t3} (note that
their \HII\ regions NGC 5, 9 and 18 correspond to NGC\,625 A, B, C, 
respectively, of {Cannon \etal\ 2003}\nocite{cannon03}). For these parameters 
we then expect a \begin{math}\frac{j_{H\alpha}}{j_{1384}}\end{math} ratio 
between  (7.15-8.04)$\times$10$^{-10}$\,erg\,sec$^{-1}$\,cm$^{-2}$\,Jy$^{-1}$,
for purely thermal emission in the presence of no reddening.

\placetable{t3}

Combining the measured \HII\ region flux densities from the radio continuum 
with the measured flux of the \HII\ regions based on our previous HST 
narrowband photometry, we calculate the 
\begin{math}\frac{j_{H\alpha}}{j_{1384}}\end{math} ratio for each of the three
most luminous \HII\ regions and use these values to derive the implied 
A$_{H\alpha}$. The low \begin{math}\frac{j_{H\alpha}}{j_{1384}}\end{math} 
values for regions (A$+$D) and C suggest that either appreciable amounts of 
extinction exist toward these starburst regions, or that the contribution of 
non-thermal sources to the radio continuum luminosity is significant.  
It is very common for A$_{H\alpha}$ calculated from the ratio of Balmer 
fluxes to radio continuum fluxes to be much larger than A$_{H\alpha,Balmer}$, 
even in the absence of nonthermal radio continuum emission (cf. {Israel \& 
Kennicutt 1980}\nocite{israel80}; {Caplan \& Deharveng 1986}\nocite{caplan86}; 
{Skillman \& Israel 1986}\nocite{skillman88c}; {Bell \etal\ 
2002}\nocite{bell02a}).  This arises simply because in non-homogeneous 
geometries A$_{H\alpha, Balmer}$ can saturate at values as low as 0.4 so that 
A$_{H\alpha, Balmer}$ frequently is only a lower limit to the true extinction.
Similar results were found by {Beck, Turner, \& Kovo (2000)}\nocite{beck00} 
for a sample of Wolf-Rayet galaxies. As shown in Table\,~\ref{t3}, the 
extinction varies widely from one \HII\ region to the next, with region B 
suffering no extinction as measured from the 
\begin{math}\frac{j_{H\alpha}}{j_{1384}}\end{math} ratio. If the low ratios 
for regions (A$+$D) and C are caused by the presence of additional extinction, 
then these results are consistent with results from both our HST/WFPC2 imaging 
and ground-based spectroscopy, where strong and variable extinction was found 
throughout the galaxy.  In the latter study, differential extinction precluded 
a detailed temporal resolution of the recent star formation history.

Unfortunately, with these data we only have one frequency with which to probe 
the radio continuum, precluding detailed spectral index work to reveal the 
contribution of potential non-thermal sources. Multi-frequency radio continuum 
observations of this system are underway (Cannon \etal\ 2004, in preparation), 
as they will help to further probe the nature of, and extinction toward, the 
\HII\ regions in this active dwarf galaxy.  We can gain some insight into this 
issue, however, by using the optical spectroscopy of \citet{skillman03b}.  The 
c(\hbeta) values calculated therein are converted to A$_V$ and shown in 
Table~\ref{t3}.  The difference in implied extinctions between the optical 
(i.e., \halpha/\hbeta) and the radio (i.e., \halpha/radio continuum) suggests 
that the nonthermal contribution is large; however, multi-frequency data will 
answer this question definitively.  With the current data, we conclude that 
the measured radio continuum luminosities, when compared to the observed 
Balmer line fluxes, are consistent with either heavy (and spatially variant) 
extinction toward some of the starburst regions (A$_{H\alpha} \sim$ 1-2\,mag), 
or the presence of a large non-thermal component of the radio continuum 
luminosity.

	
\section{Conclusions}
\label{S5}

We have presented new ATCA multi-configuration \HI\ synthesis imaging of the
nearby dwarf starburst galaxy NGC\,625.  The \HI\ morphology appears  
smooth on global scales; low-resolution column density images show one 
strong central \HI\ peak surrounded by a moderate-size low-density halo.  
However, examining the morphology at higher resolution reveals multiple peaks 
in the column density, suggesting that the dynamics of the neutral gas are 
quite complex.

Examination of the velocity field shows a complicated structure, with \HI\ 
emission detected near the disk across the full velocity range. We find 
velocity structure in the \HI\ data cubes that appears consistent with the 
signature of neutral gas blowout from the main disk.  While at our resolution 
we are not sensitive to small-scale structure (which may show holes and other 
signatures of the influence of massive stars), we are sensitive to the global 
effects that this star formation is having on the ISM of NGC\,625. Because we 
do not detect large \HI\ holes, and given the results of our HST/WFPC2 star 
formation history analysis \citep{cannon03}, we prefer the interpretation 
that the current outflow episode is not the result of one major star 
formation region, but rather the collective influence of the extended (both 
temporally and spatially) star formation event which NGC\,625 has undergone.   

Active blowout is the only interpretation that is capable of reproducing all 
of the characteristics of the position-velocity diagrams along the major 
axis of the system.  Other potential explanations of the complex velocity 
structure, including superposed disks, counter-rotating disks, infall, 
ongoing assimilation, high-velocity clouds, and ISM turbulence, are discussed 
and rejected, each failing to reproduce all of the observed velocity 
structure.  While a detailed hydrodynamical simulation is beyond the scope of 
this paper, we encourage further such investigations to better understand the 
mechanisms involved in creating blowout in low-mass systems. 

At the current sensitivity, we find no obvious external trigger for the 
current burst of star formation.  As NGC\,625 is a comparatively isolated 
galaxy, we thus conclude that it has been able to sustain a heightened 
star formation rate due to the relatively large reservoir of high-column 
density gas in the central disk regions.  The initial triggering mechanism for
this extended burst remains unknown.  

We detect strong radio continuum emission from the three largest star formation
regions in NGC\,625.  Comparing to our HST/WFPC2 \halpha\ fluxes, we find 
that either appreciable amounts of extinction exist toward these regions, or 
that there is a strong nonthermal component to the radio continuum emission. We
have undertaken multi-frequency continuum observations of this system to 
further differentiate between these two scenarios.  

NGC\,625 is a remarkable system.  It appears to be one of only a few 
dwarf galaxies that have shown neutral gas blowout from the active star 
formation regions.  It is the only dwarf galaxy known to show the kinematic 
signature of blowout, but not to show a large evacuated region near the active
star formation region (although such a hole(s) may have gone undetected in the 
present experiment, at our resolution).  Given the detection of diffuse soft 
x-ray gas above the plane of this system \citep{bomans98}, it appears that the 
extended massive star formation event in NGC\,625 \citep{cannon03} has 
achieved a similar integrated effect.  This has important implications for the 
evolution of dwarf galaxies.  Given the correct conditions, they are able to 
sustain extended star formation events, which will have dramatic effects on 
the ISM and surrounding IGM.

\acknowledgements

The authors appreciate the insightful comments of an anonymous referee that 
helped to improve this manuscript.  J.\,M.\,C. is supported by NASA Graduate 
Student Researchers Program (GSRP) Fellowship NGT 5-50346, and is grateful to 
the National Radio Astronomy Observatory Foreign Telescope Travel Fund for 
international travel support.  E.\,D.\,S. is grateful for partial support from 
NASA LTSARP grant NAG5-9221 and the University of Minnesota.  This 
research has made use of: the NASA/IPAC Extragalactic Database (NED) which is 
operated by the Jet Propulsion Laboratory, California Institute of Technology, 
under contract with the National Aeronautics and Space Administration; and 
NASA's Astrophysics Data System. 

\clearpage


\clearpage

\clearpage

\clearpage

\clearpage

\clearpage
\thispagestyle{empty}

\clearpage

\clearpage
\begin{deluxetable}{ccc}
\tabletypesize{\scriptsize}
\tablecaption{Basic Parameters of NGC\,625}
\tablewidth{0pt}
\tablehead{
\colhead{Property}         
&\colhead{Value} 
&\colhead{Reference}}
\startdata
Distance (Mpc)  	&3.89$\pm$0.22		&\citet{cannon03}\\
M$_B$			&$-$16.3		&\citet{marlowe97}\\
Gal. Lat. (\degree)	&$-$73.1		&$--$\\
Foreground A$_V$ (mag.)		&0.05			&\citet*{schlegel98}\\
12\,$+$\,log(O/H) 	&8.14$\pm$0.02		&\citet{skillman03b}\\
Current SFR (\msun\,yr$^{-1}$)  &0.05  		&\citet{skillman03a}\\
\HI\ Mass (10$^8$ \msun) &1.1$\pm$0.2		&This work\\
V$_{Helio}$ (\kms)	&413$\pm$5 		&This work\\
V$_{20}$ (\kms)		&95$\pm$2		&This work\\
V$_{50}$ (\kms)		&62$\pm$2		&This work\\
\enddata
\label{t1}
\end{deluxetable}

\clearpage
\begin{deluxetable}{ccc}
\tabletypesize{\scriptsize}
\tablecaption{ATCA Observations of NGC\,625}
\tablewidth{0pt}
\tablehead{
\colhead{Array}         
&\colhead{Observation} 
&\colhead{T$_{INT}$}\\
\colhead{Configuration} 
&\colhead{Date(s)}        
&\colhead{(Minutes)}}
\startdata
6 F	&2001, 27 May 			&584\\
375	&2001, 14, 16 Jun	 	&176\\
1.5 A   &2001, 15 Aug	 		&579\\
6 B	&2001, 19, 20 Aug		&716\\
750 D	&2001, 23 Sep		    	&596\\
EW 352	&2001, 11, 12, 13, 15, 17 Oct	&1263\\
6 D     &2001, 25 Nov, 4 Dec 		&606\\
\enddata
\label{t2}
\end{deluxetable}

\clearpage
\begin{deluxetable}{cccccc}
\tabletypesize{\scriptsize}
\tablecaption{Radio Continuum vs. \halpha\ Comparison}
\tablewidth{0pt}
\tablehead{
\colhead{\HII\ Region(s)\tablenotemark{a}}         
&\colhead{HST \halpha\ Flux\tablenotemark{b}}
&\colhead{T$_e$\tablenotemark{c}} 
&\colhead{1384\,MHz Flux}
&\colhead{Implied A$_{H\alpha}$\tablenotemark{d}} 
&\colhead{Optical A$_{H\alpha, Balmer}$\tablenotemark{e}}\\ 
\colhead{}
&\colhead{(10$^{-14}$\,\ergseccm)}
&\colhead{(K)}
&\colhead{Density (mJy)}
&\colhead{(mag)}
&\colhead{(mag)}}
\startdata
\vspace{0.1 cm}
NGC\,625\,A$+$D	&211\,$\pm$\,10	&10900$^{+115}_{-109}$  &7.5\,$\pm$\,1.5    
&1.2\,$\pm$\,0.5 &0.27\,$\pm$\,0.07\\ 	
\vspace{0.1 cm}
NGC\,625\,B	&112\,$\pm$\,6	&10460$^{+213}_{-193}$  &1.2\,$\pm$\,0.24  
&$-$0.2\,$\pm$\,0.2               &0.02\,$\pm$\,0.09\\
NGC\,625\,C     &7.8\,$\pm$\,0.4 &12810$^{+460}_{-395}$	&0.86\,$\pm$\,0.17  
&2.2\,$\pm$\,0.5 &0.48\,$\pm$\,0.09\\
\enddata
\label{t3}
\tablenotetext{a}{Adopting the nomenclature of 
\citet{cannon03}; see Figure~\ref{figcap6}.}
\tablenotetext{b}{Drawn from \citet{cannon03}}
\tablenotetext{c}{Drawn from \citet{skillman03b}}
\tablenotetext{d}{Calculated using \halpha\ values from \citet{cannon03}
and the radio continuum values derived in this work.}
\tablenotetext{e}{Applying spectroscopic c(\hbeta) values from 
\citet{skillman03b}, and the nomenclature of \citet{bell02a}.}
\end{deluxetable}
\end{document}